\documentclass[sigconf]{acmart}
\AtBeginDocument{%
  }

\setcopyright{acmlicensed}
\copyrightyear{2026}
\acmYear{2026}
\acmDOI{10.1145/XXXXXXX.XXXXXXX}

\acmConference[CHI '26]{CHI Conference on Human Factors in Computing Systems}{April 13--17, 2026}{Barcelona, Spain}

\acmISBN{979-8-4007-XXXX-X/26/04}

\usepackage{tabularx}
\usepackage{booktabs}
\usepackage{ragged2e}
\usepackage{enumitem}

\begin{document}

\title{Prompting Destiny: Negotiating Socialization and Growth in an LLM-Mediated Speculative Gameworld}

\author{Mandi Yang}
\email{mandy1203@mail.nankai.edu.cn}
\affiliation{%
  \institution{Nankai University}
  \city{Tianjin}
  \country{China}
}

\author{Zhiqi Gao}
\affiliation{\institution{The Chinese University of Hong Kong, Shenzhen} \city{Shenzhen} \state{Guangdong}\country{China}}
\email{zhiqigao@link.cuhk.edu.cn}

\author{Yibo Meng}
\affiliation{%
  \institution{Tsinghua University}
  \city{Beijing}
  \country{China}
}

\author{Dongyijie Primo Pan}
\authornote{Corresponding author.}
\email{dpan750@connect.hkust-gz.edu.cn}
\affiliation{%
  \institution{The Hong Kong University of Science and Technology (Guangzhou)}
  \city{Guangzhou}
  \state{Guangdong}
  \country{China}
}

\renewcommand{\shortauthors}{Yang et al.}

\begin{abstract}
We present an LLM-mediated role-playing game that supports reflection on socialization, moral responsibility, and educational role positioning. Grounded in socialization theory, the game follows a four-season structure in which players guide a child-prince through morally charged situations and compare the LLM-mediated NPC’s differentiated responses across stages, helping them reason about how educational guidance shifts with socialization. To approximate real educational contexts and reduce score-chasing, the system hides real-time evaluative scores and provides delayed, end-of-stage “growth” feedback as reflective prompts. We conducted a user study (N=12) with gameplay logs and post-game interviews, analyzed via reflexive thematic analysis. Findings show how players negotiated responsibility and role positioning, and reveal an entry-load tension between open-ended expression and sustained engagement. We contribute design knowledge on translating sociological models of socialization into reflective AI-mediated game systems.
\end{abstract}



\maketitle

\section{INTRODUCTION}

Education is not simply the transmission of knowledge, but a core mechanism of socialization through which values, norms, roles, and moral orientations are gradually internalized via relationships, authority, and responsibility in everyday life~\cite{berger1966social,mead1934mind,Goffman1959}. Yet socialization is often \emph{invisible}: the consequences of educational guidance rarely become legible in the moment and may only surface through conflict, resistance, or emotional escalation. Moreover, socialization is stage-dependent---what counts as authority, care, and responsibility can shift across developmental phases.

This raises a structural challenge: in everyday life, educational relationships are difficult to isolate as observable objects, leaving people limited opportunities to reflect, in a structured way, on how they shape others and are shaped in return. Interactive systems---especially role-playing games---offer a promising route by placing players in structured situations where relational authority, responsibility allocation, and moral consequences can be \emph{experienced} and made reflectable, turning play into a kind of social laboratory~\cite{sengers2005reflective,flanagan_nissenbaum_2014}. With large language models (LLMs), this potential expands: LLM-mediated NPCs can generate socially interpretable, context-sensitive responses during interaction, making social relations, meaning, and responses dynamically produced rather than pre-scripted~\cite{park2023generativeagentsinteractivesimulacra}.

However, most \textbf{raising simulation games}~\cite{pasin2011impact} operationalize ``education'' through visible stat progression and real-time feedback (e.g., \emph{Volcano Princess})~\cite{VolcanoPrincessSteam}, which can invite metric chasing and reduce socialization to performance optimization~\cite{Mogavi2022GamificationSpoils}. Meanwhile, open-ended LLM role-play can enable expressive freedom but often lacks a stable, theory-grounded scaffold for understanding how educational guidance shifts across stages of socialization. We address this gap by translating a staged socialization model into a four-season interaction structure, and by withholding real-time evaluative scores (an \emph{anti-visualization} strategy) in favor of delayed, end-of-stage reflective prompts. We use \emph{entry load} to denote the cognitive and linguistic burden introduced by open-ended input.

\textbf{Therefore, we pose the following research question:}
\begin{itemize}[topsep=0pt,itemsep=0pt,parsep=0pt]
  \item \textbf{RQ:} How can we design stage-based AI-NPC interactions and feedback timing in an LLM-mediated role-playing game to support reflection on educational role positioning and responsibility across socialization stages?
\end{itemize}

To explore this question, we designed an LLM-mediated role-playing game grounded in socialization theory. The game maps a four-stage socialization model onto a four-season narrative arc, where players guide a child-prince through morally charged situations and reflect on how LLM-mediated NPCs respond differently across stages. To approximate real-life educational contexts and reduce score-chasing, the system withholds real-time evaluative scores and provides delayed end-of-stage ``growth'' summaries as reflective prompts. We conducted a user study (N=12) with gameplay logs and post-game interviews, analyzed using reflexive thematic analysis, and report how players engaged with educational role positioning and responsibility across stages, alongside a tension between open-ended expression and entry load.

This work contributes design knowledge for translating sociological models of socialization into reflective AI-mediated role-playing systems, and implications for designing AI-NPC interactions that make educational relationships and their consequences more legible.\vspace{-0.6em}

\section{RELATED WORK}

Sociological studies of tabletop role-playing describe RPGs as negotiated social worlds grounded in role-making, shared norms, and accountability in interaction~\cite{Goffman1959,fine2002shared}. In HCI, reflective and values-oriented game research shows that interactive experiences can be structured to surface moral tension and support reflection beyond moment-to-moment performance~\cite{sengers2005reflective,flanagan_nissenbaum_2014,SCHRIER2023221,10.1145/3629606.3629618}. Recent work demonstrates that LLMs are increasingly embedded in interactive narrative systems, enabling real-time authoring, emergent storytelling, and agent-like NPC interactions~\cite{10.1145/3613904.3642579,10.1145/3706599.3720212}. Studies further document both the promise and limitations of LLM-driven NPC dialogue, including open-ended agency as well as issues of unpredictability and coherence breakdown~\cite{Peng2024Player-Driven,Christiansen2024Exploring}. Beyond games, HCI research shows that role-play and playful probing can support reflection on LLM agency~\cite{10.1145/3678884.3681912,10.1145/3711015}, and that long-term systems can deliver sustained reflective feedback over time~\cite{10.1145/3757616}. Framing LLM interaction as ``role play'' has also been proposed as a way to discuss agent-like behavior without collapsing into anthropomorphism~\cite{shanahan2023role}. However, we still lack design knowledge about LLM-mediated role-play systems that translate staged socialization theory into interaction structures for reflecting on \emph{educational role positioning} and responsibility, particularly regarding how stage boundaries and feedback timing shape reflection. We address this gap by introducing a staged socialization arc with delayed “growth” feedback (anti-visualization), and by examining how players interpret educational responsibility under the trade-offs introduced by entry load.

\section{METHODS}

\subsection{Narrative and Interaction Design}
Our system uses an LLM-driven narrative engine (GPT-4) to generate branching plots in a non-linear story space. At key decision points, players provide structured choices and/or free-text inputs to express intentions, moral stances, and guidance strategies. Conditioned on the evolving narrative context, the engine updates story state, emotional tone, and character relationships, enabling both within-session branching and longer arcs with accumulated consequences.

The experience follows a four-stage (seasonal) socialization arc: Spring (Initiation), Summer (Exploration), Autumn (Consolidation), and Winter (Consequences). Early stages introduce norms and relationships, while later stages escalate ambiguity and moral tension and surface cumulative outcomes. Within each stage, we embed interaction nodes such as decision trees, free-text justifications, and (in the multi-user version) group voting. Nodes are tagged with narrative context identifiers so the LLM can produce coherent, context-aware updates (e.g., a famine decision in Autumn affects outcomes in Winter). Input affordances are varied across stages and contexts to scaffold reflection while managing entry load.

\subsection{LLM-Based Feedback and System Implementation}
Consequence feedback is generated through a three-channel loop: (1) narrative events (e.g., unrest), (2) character-level responses (e.g., trust changes), and (3) environmental updates (e.g., resource shifts). To discourage score-chasing, evaluative signals are maintained internally and surfaced primarily through narrative consequences and end-of-stage summaries rather than real-time evaluative scores, foregrounding moral accountability as experienced outcomes.

We implemented a multiplayer-capable prototype in Unity 3D with Photon Unity Networking. Python orchestration scripts manage sessions and interface with the LLM API. The interface (Fig.~1) presents narrative prompts, chat logs, and role-specific dashboards, while gameplay logs capture decisions and interactions for post-hoc reflection and analysis. The modular architecture supports rapid iteration on scenarios, roles, and rule sets.

\begin{figure*}[t]
  \centering
  \includegraphics[width=\textwidth]{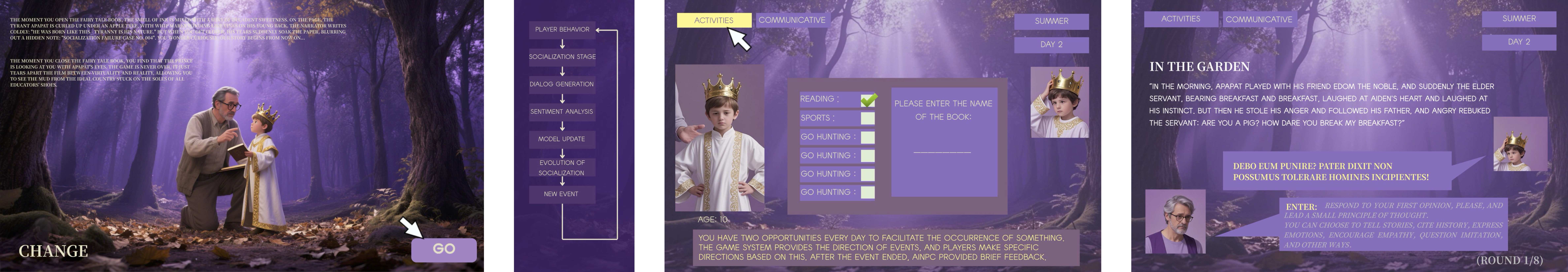}
  \caption{Screenshots of the game interface illustrating key moments in the staged role-play experience. From left to right: the opening scene that sets the narrative context; a control panel for stage progression and player actions; an activity selection view where players choose options for the prince; and an in-game dialogue scene with the NPC mentor and the prince.}
  \Description{A four-panel screenshot collage of a purple-themed narrative role-play game UI. Panel 1 shows an opening scene in a forest with a mentor facing a young prince and a large “Go” button. Panel 2 shows a vertical control menu listing gameplay stages and options. Panel 3 shows an activity selection screen with checkboxes and a text input area. Panel 4 shows a dialogue scene titled “In the Garden,” with character portraits (mentor and prince) and a text box for player input.}
  \label{fig:ui}
\end{figure*}

\subsection{User Study}

\subsubsection{Participants and Recruitment}
We recruited 12 participants (7 female, 5 male), aged 21 to 32, with diverse gaming experience and sociocultural backgrounds. Detailed participant demographics are provided in Appendix~C. Participants were recruited via snowball sampling. Each participant received a \$7-equivalent gift as compensation. The study protocol was approved by our institution's IRB.

\subsubsection{Procedure}
The study consisted of three stages, each lasting approximately 10--15 minutes. First, participants completed a full gameplay session using our custom game interface. While the system is a multiplayer-capable prototype, this preliminary study evaluated the experience in individual sessions. The narrative and decision points were generated in real time by an integrated LLM; participants made choices at key junctures based on story context and character interactions, which influenced plot branching. Throughout gameplay, an Evaluation LLM ran in the background to generate internal interpretive signals that supported staged feedback; these signals were not shown to participants to avoid influencing moral reasoning and to reduce score-chasing effects. Second, upon story completion, the system presented a stage-based growth feedback summary across four dimensions as post-hoc feedback (used as a reflection prompt in the interview). Third, we conducted a semi-structured interview (10--15 minutes) to probe participants' behavioral responses, shifts in social cognition, and reflections on educational control, responsibility, and self--other understanding. We used gameplay logs to contextualize and triangulate interview accounts.

\subsection{Analysis}
We report qualitative findings derived from reflexive thematic analysis (RTA) of post-game interviews and gameplay logs, following the six-phase process commonly associated with thematic analysis~\cite{vaismoradi2013content,braun2019reflecting}. Analysis was iterative and interpretive: we generated initial codes, developed and refined candidate themes through reflexive memoing and discussion, and revisited transcripts/log excerpts to ensure themes captured patterned meanings across participants. Core interview questions included: (1) ``How did you understand your role in the Prince's growth?'' (2) ``Were there any events that made you question the controllability of education?'' and (3) ``What kind of relationship do you feel you formed with the Prince during the game?''

The Evaluation LLM additionally produced stage-based growth feedback summaries from gameplay traces. We treat these summaries as \emph{system-generated reflective artifacts} (not validated psychometric measures) and use them to support participants' recall and reflection during interviews. To reduce output variability, we fixed the Evaluation LLM rubric and prompting template across sessions and interpret its outputs as reflective prompts rather than ground-truth measurements. For transparency, we include two supplementary radar visualizations (group-level and a representative individual example) in Appendix~B as descriptive reference.

\section{FINDINGS}
Our reflexive thematic analysis (RTA) generated five cross-stage themes that capture how participants made sense of educational role positioning, responsibility, and moral consequences as the four-season arc unfolded. Rather than optimizing for ``winning,'' participants described shifts toward interpreting consequences, re-positioning responsibility, and re-framing their relationship with the Prince. Two supplementary radar visualizations of system-generated growth feedback are provided in Appendix~B for descriptive reference.

\textbf{Uncontrollable Educational Impact.}
Across stages, participants frequently described mismatches between intention and outcome: actions meant as ``guidance'' sometimes amplified harm or produced unexpected social consequences. These moments disrupted a sense of educational control and prompted re-attribution of responsibility under uncertainty. As one participant reflected, ``I didn't expect a story to turn him into a violent teenager'' (P09).

\textbf{Social Role Shift.}
Participants rarely remained stable ``mentors.'' When persuasion failed or the Prince resisted, they reported shifting into negotiators, mediators, or stepping back from guidance altogether, treating education as a relational process rather than a one-way intervention. As P11 put it, ``I tried to make him obedient, but he didn't want to be educated at all.''

\textbf{Moral Situational Tension.}
Dilemmas frequently required moral stance-taking without clear ``correct'' options, often framed as conflicts among stakeholders with competing claims. Participants described deliberating not only outcomes but also the legitimacy of their own authority to decide. ``I'm not sure who to support---a prince or a villager'' (P05).

\textbf{AI Cognitive Transformation.}
Over time, some participants re-framed the Prince from a reactive object into an emotionally salient other. This shift redirected attention from instrumental success toward relational consequences, such as trust, fear, or care within the storyworld. As P01 noted, ``I began to worry about the prince's feelings, not how to win.''

\textbf{Self-Reflection Upgrade.}
Finally, participants used narrative outcomes to reflect on their own dispositions as ``educators,'' sometimes drawing parallels between their guidance practices and the authority structures encountered in the storyworld. This reflexive comparison surfaced discomfort and prompted re-evaluation of one’s role identity: ``In fact, the way I educate him is no different from his father'' (P05).

\begin{figure*}[t]
  \centering
  \includegraphics[width=\linewidth]{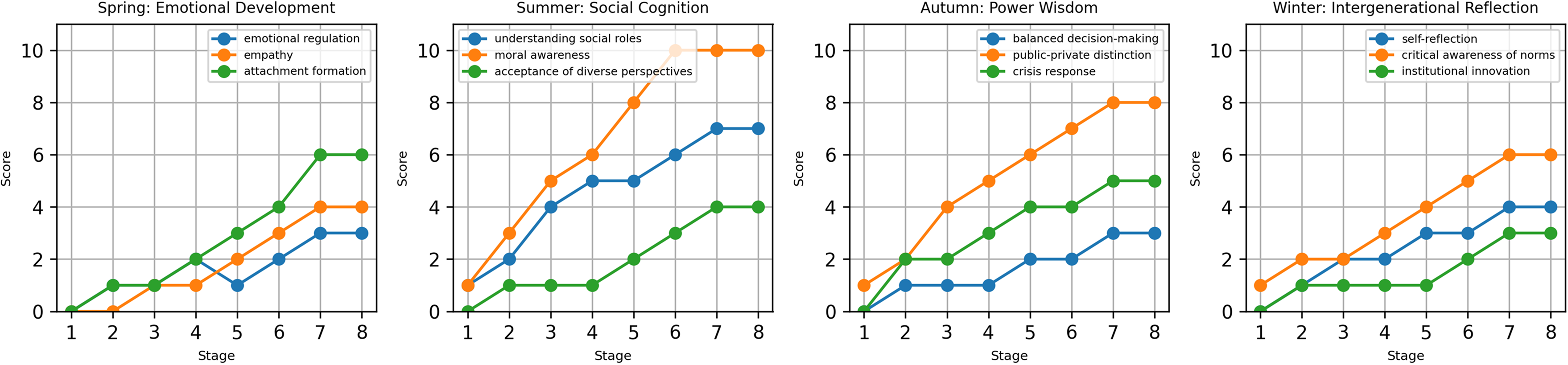}
  \caption{An illustrative staged growth feedback summary generated by the system for one participant at the end of play. We use this artifact to prompt reflection in interviews; it is not treated as a validated psychometric measure.}
  \label{fig:feedback_artifact}
  \Description{A multi-panel line chart showing four dimensions of a character growth summary across staged gameplay, used as post-hoc feedback for reflection rather than real-time scoring.}
\end{figure*}

Beyond these themes, participants highlighted the delayed (anti-visualization) feedback design as a boundary marker for reflection. By withholding in-the-moment scores, the system discouraged score-chasing and instead prompted moments of ``sudden realization'' at stage transitions about earlier choices and everyday interactions they had overlooked. Figure~\ref{fig:feedback_artifact} shows one representative staged growth feedback summary used as a reflection prompt; it is not treated as a validated psychometric measure.

\section{DISCUSSION}

\subsection{Stage-Based Socialization and Shifts in Educational Role Positioning}
Across the four-season arc, participants did not experience ``education'' as isolated choices, but as a relational process in which authority, care, and responsibility were continuously re-positioned in interaction. The shift we observed---from attempting directive control to negotiating guidance within a relationship---aligns with sociological accounts of socialization as the gradual internalization of norms and identities through everyday encounters and role performances~\cite{mead1934mind,Goffman1959,berger1966social}. The seasonal structure functioned as a temporal scaffold that made this process experientially legible: early stages encouraged participants to adopt an educator-like stance, while later stages foregrounded resistance, ambiguity, and consequences, prompting reflection on how educational authority can fail, backfire, or become ethically contested. This supports prior HCI arguments that reflective play can surface values and moral tension as objects of interpretation rather than optimization~\cite{sengers2005reflective,flanagan_nissenbaum_2014,CHI24}.

\subsection{Delayed Feedback as Modeling Education as a Cumulative Social Process}
Participants described the end-of-stage ``growth'' summaries as moments that re-framed earlier interactions, not as scores to calibrate in real time. This suggests a design rationale beyond ``anti-score-chasing'': withholding real-time evaluative scores models how educational consequences are often cumulative and only become visible retrospectively through conflict, resistance, or emotional rupture. In contrast to growth systems that invite metric chasing (common in child-raising games), delayed feedback can interrupt performance logic and support sensemaking about responsibility and moral consequences~\cite{Mogavi2022GamificationSpoils}. This is consistent with prior work on reflection and feedback timing, where stage boundaries can serve as prompts for retrospective interpretation rather than continuous numerical calibration~\cite{bentvelzen2022revisitingreflection,jane2024timingfeedback}.

At the same time, participants' reports point to two practical design tensions. First, open-ended input increased \emph{entry load} in later stages, suggesting the need for lightweight scaffolds (e.g., justification templates, segmented prompts, or contextual recall cues) to preserve agency while reducing cognitive and linguistic burden~\cite{sweller1988cognitiveload}. Second, because the Evaluation LLM produced system-generated summaries, we treat these outputs as reflective artifacts rather than validated measures, and recommend transparent rubrics and conservative interpretation given known variability in LLM behavior~\cite{10.1145/3290605.3300233,10.1145/3442188.3445922}.

\subsection{Limitations and Scope Conditions}
This work is an exploratory study with a small sample (N=12) and snowball recruitment, which may introduce self-selection bias. The study evaluated a multiplayer-capable prototype in individual sessions; future work should examine multi-user dynamics directly. We did not include a control condition (e.g., a non-LLM or non-delayed-feedback variant), so we do not make causal claims about the effects of any single mechanism. Our findings are most applicable to staged, narrative role-play settings where educational role positioning and responsibility formation are central; transfer to other genres, cultures, or less structured AI-mediated role-play experiences requires further investigation.

\section{Conclusion}
We introduced an LLM-mediated role-play game that translates a staged socialization arc into a four-season structure with delayed (anti-visualization) feedback to support reflection on educational role positioning, responsibility, and moral consequences. Through RTA of interviews and gameplay logs, we surfaced five cross-stage themes and a key design tension---entry load---that highlights trade-offs between open expression and accessible reflection. We hope these findings inform future designs of reflective play experiences with generative AI.

\section*{Generative AI Use Disclosure}
We used LLMs in two ways: within the research system and as a development aid. GPT-4 generated narrative and NPC dialogue, and an Evaluation LLM produced stage-based growth summaries, which we treat as reflective artifacts rather than validated psychometric measures; the evaluation rubric and prompts were fixed across sessions. 

For development, we used Cursor (GPT-4o) for implementation support, with all outputs reviewed by the authors. No raw interview transcripts or identifiable participant data were uploaded to external LLM services. All analysis and claims are the responsibility of the authors.

\bibliographystyle{ACM-Reference-Format}
\bibliography{sample-base}

\newpage
\appendix

\onecolumn
\section*{Appendix}
\section{Game Prompt}

We are about to engage in a social simulation game that models the growth of a ruler and the development of society. Players will assume the role of the ruler’s mentor, applying different guidance strategies and developmental policies at various stages.

\subsection{Narrative Background}

The player, a sociologist, finds themselves transported into the fictional world of \textit{The Tyrannical King Apapat}. Their mission is to alter the tragic fate of the king’s brutal reign. The game is divided into four seasonal chapters—Spring, Summer, Autumn, and Winter—each corresponding to a specific socialization phase:

\begin{itemize}
    \item \textbf{Spring (Childhood Socialization):} Legitimacy construction through internalization of basic norms and values.
    \item \textbf{Summer (Anticipatory Socialization):} Learning and preparing for future social roles.
    \item \textbf{Autumn (Developmental Socialization):} Adapting to new environmental demands.
    \item \textbf{Winter (Reverse Socialization):} Influence from younger generations on elders’ cognition; focus on legacy planning and power transition.
\end{itemize}

The player receives all initial character settings and is tasked with simulating the game’s progression accordingly.

\subsection{Character Profiles}

\paragraph{Prince Apapat}
\begin{itemize}
    \item \textbf{Age:} 7
    \item \textbf{Personality:} Sensitive, curious, emotionally volatile
    \item \textbf{Location:} Central area of Kapas Palace (e.g., Throne Room, Study, Garden)
    \item \textbf{Profile:} Apapat is innocent, highly impressionable, and curious. The player’s decisions will shape his values and personality.
\end{itemize}

\paragraph{King Kalman}
\begin{itemize}
    \item \textbf{Age:} 35
    \item \textbf{Personality:} Resolute, authoritarian, controlling
    \item \textbf{Location:} Throne Room and Training Ground
    \item \textbf{Profile:} Kalman values power and discipline. The player may influence him to reflect on his harsh parenting.
\end{itemize}

\paragraph{Queen Eliza}
\begin{itemize}
    \item \textbf{Age:} 32
    \item \textbf{Personality:} Resilient, responsible, decisive
    \item \textbf{Location:} Training Ground and Council Chamber
    \item \textbf{Profile:} Eliza supports gentle education but often appears emotionally distant due to her duties.
\end{itemize}

\paragraph{Advisor Luca}
\begin{itemize}
    \item \textbf{Age:} 40
    \item \textbf{Personality:} Shrewd, pragmatic, diplomatic
    \item \textbf{Location:} Study and Archive Room
    \item \textbf{Profile:} A strategic thinker and neutral mediator. Can be an ally depending on player guidance.
\end{itemize}

\paragraph{Court Knight Ryan}
\begin{itemize}
    \item \textbf{Age:} 28
    \item \textbf{Personality:} Loyal, brave, stubborn
    \item \textbf{Location:} Training Ground and Fountain Square
    \item \textbf{Profile:} A discipline advocate who may soften his approach through player influence.
\end{itemize}

\paragraph{Maid Vera}
\begin{itemize}
    \item \textbf{Age:} 24
    \item \textbf{Personality:} Cheerful, witty, empathetic
    \item \textbf{Location:} Garden and Dining Hall
    \item \textbf{Profile:} Uses play and storytelling to educate; avoids conflict due to her status.
\end{itemize}

\paragraph{Young Noble Aiden}
\begin{itemize}
    \item \textbf{Age:} 9
    \item \textbf{Personality:} Playful, curious, rebellious
    \item \textbf{Location:} Garden and Fountain Square
    \item \textbf{Profile:} Engages Apapat in games and mischief; promotes empathy through friendship.
\end{itemize}

\paragraph{Scholar Celine}
\begin{itemize}
    \item \textbf{Age:} 38
    \item \textbf{Personality:} Patient, rigorous, intellectual
    \item \textbf{Location:} Study and Archive Room
    \item \textbf{Profile:} Fosters wisdom through logic, history, and astronomy.
\end{itemize}

\paragraph{Captain Gale}
\begin{itemize}
    \item \textbf{Age:} 30
    \item \textbf{Personality:} Loyal, disciplined, serious
    \item \textbf{Location:} Training Ground and Fountain Square
    \item \textbf{Profile:} Focused on physical training and collective responsibility.
\end{itemize}

\paragraph{Musician Nora}
\begin{itemize}
    \item \textbf{Age:} 26
    \item \textbf{Personality:} Expressive, emotional, creative
    \item \textbf{Location:} Garden and Dining Hall
    \item \textbf{Profile:} Encourages emotional development through music and poetry.
\end{itemize}

\subsection{Random Events}

Each season may feature 5–8 random events. Most events are initiated by Prince Apapat or connected to him.

\subsubsection{Scene Structure}

\begin{enumerate}
    \item \textbf{Scene Background:}  
    Events must occur in palace-appropriate settings and align with the narrative and socialization themes.

    \item \textbf{Initiator:}  
    Typically Prince Apapat, though other related characters may initiate events. Initiators adapt their memory based on player dialogue.

    \item \textbf{Dialogue Strategy:}  
    Each event consists of a dialogue scene limited to 8 dialogue turns. You will conclude each scene.

    \item \textbf{Decision Assessment:}  
    After each scene, assess and provide feedback on the character’s state based on the player’s input. Update memory and adjust numerical attributes accordingly.

    \begin{quote}
        \textit{Example: “The prince felt very happy and understood that beating servants is wrong. He left.”}  
        *(Evaluation: Player response was persuasive and cited historical reasoning. Brutality index reduced significantly.)*
    \end{quote}

    \item \textbf{Decision Options:}  
    Provide four divergent decision paths per scene for the ruler to choose from.
\end{enumerate}

\section{Supplementary Radar Visualizations of System-Generated Growth Feedback}
\label{app:radar}

\begin{figure}[htbp]
  \centering
  \includegraphics[width=\linewidth]{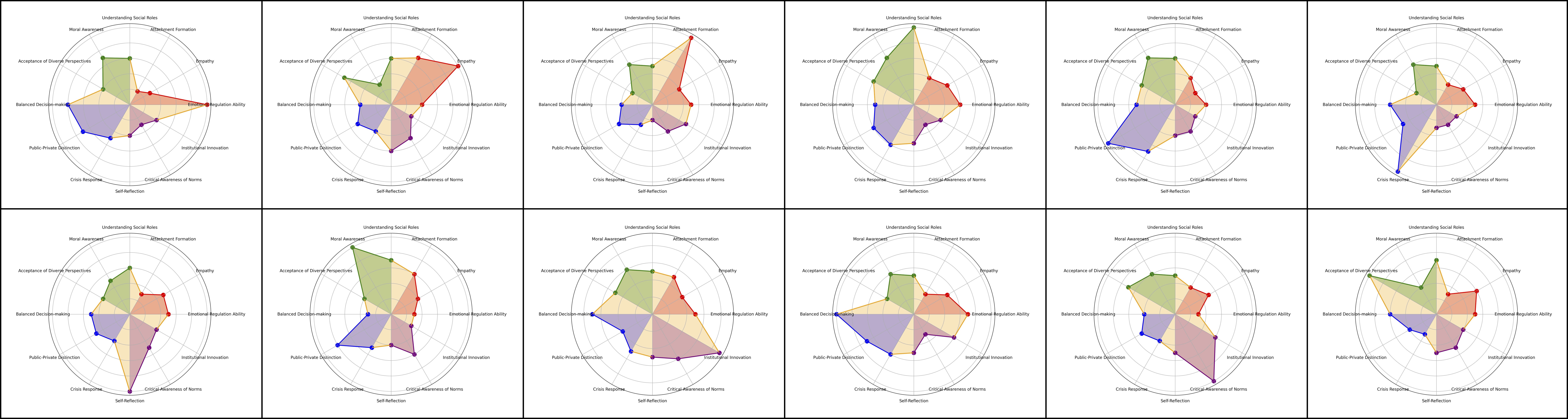}
  \caption{Supplementary radar visualization summarizing system-generated growth feedback across participants. This figure is provided for reference and is not treated as a validated psychometric measure.}
  \label{fig:radar_group}
  \Description{A set of radar charts showing multiple dimensions of a system-generated growth feedback summary across participants.}
\end{figure}

\begin{figure}[htbp]
  \centering
  \includegraphics[width=\linewidth]{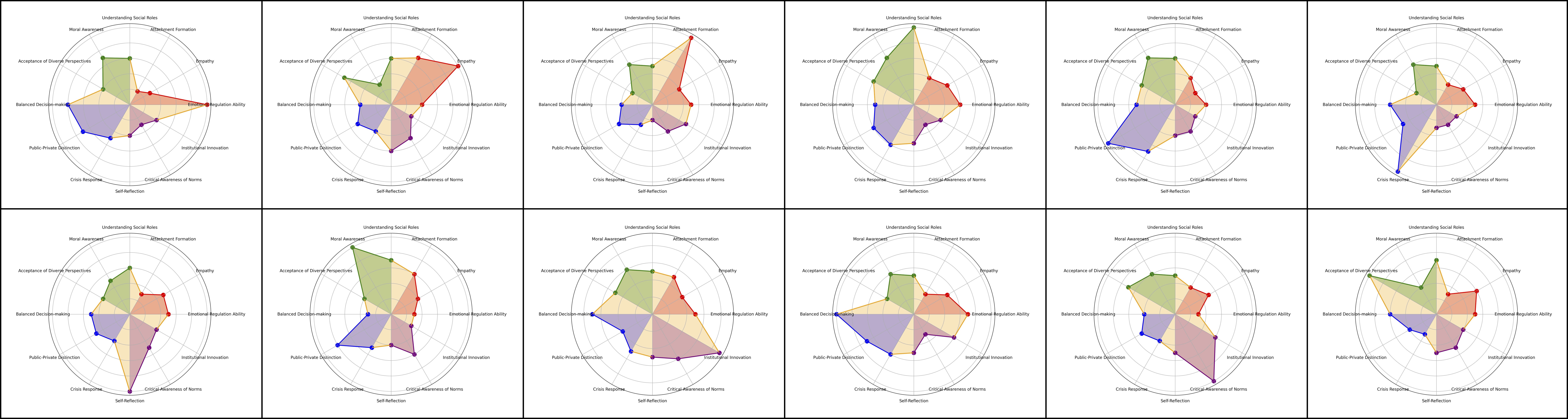}
  \caption{Supplementary radar visualization for a representative individual participant. This figure is provided for reference and is not treated as a validated psychometric measure.}
  \label{fig:radar_individual}
  \Description{A radar chart showing multiple dimensions of a system-generated growth feedback summary for one representative participant.}
\end{figure}

\section{Participant Overview}
\begin{table}[H]
  \centering
  \small
  \setlength{\tabcolsep}{6pt}
  \renewcommand{\arraystretch}{1.12}
  \caption{Participant demographics. Gaming experience was self-reported.}
  \label{tab:participants}
  \begin{tabular}{|l|l|l|l|}
    \hline
    \textbf{ID} & \textbf{Gender} & \textbf{Age} & \textbf{Gaming Experience} \\
    \hline
    P01 & Female & 22 & Moderate \\
    P02 & Male   & 24 & High \\
    P03 & Male   & 26 & Moderate \\
    P04 & Female & 21 & Low \\
    P05 & Male   & 23 & Moderate \\
    P06 & Female & 25 & High \\
    P07 & Female & 27 & Moderate \\
    P08 & Male   & 29 & High \\
    P09 & Female & 32 & Moderate \\
    P10 & Male   & 28 & Moderate \\
    P11 & Female & 24 & Low \\
    P12 & Female & 26 & Moderate \\
    \hline
  \end{tabular}
\end{table}

\end{document}